\documentstyle [12pt] {article}
\title{DARK SPRING - A SIMPLE INTERPRETATION OF THE
SUSSKIND- HOROWITZ-POLCHINSKY CORRESPONDENCE
BETWEEN SCHWARZSCHILD BLACK HOLE AND STRINGS}

\author{Vladan Pankovi\'c, Darko V. Kapor, Miodrag Krmar\\
Department of Physics, Faculty of Sciences, 21000 Novi Sad,\\ Trg
Dositeja Obradovi\'ca 4., Serbia, vpankovic@if.ns.ac.yu}

\date {}
\begin{document}
\maketitle \vspace {0.5cm}
 \vspace {0.5cm}

\begin {abstract}
In this work we suggest a simplified interpretation of
Susskind-Horowitz-Polchinski correspondence between Schwarzschild
black hole and strings. Firstly, similarly to naive, classical
mechanical Laplace determination of the Schwarzschild radius, we
suggest a simple, classical mechanical equation. It determines
amplitude of such sufficiently strong classical elastic force that
forbids escape of a Planck mass particle moving by speed of light
from end of corresponding classical elastic spring, simply called
dark spring. Also, by use of a formal identity between given
elastic force and Schwarzschild gravitational "force", we
introduce phenomenologically a simple quantization rule. It states
that circumference (corresponding to elastic force amplitude
equivalent formally to Schwarzschild radius) holds natural number
of corresponding reduced Compton's wave length. (It is deeply
analogous to Bohr's quantization postulate in Bohr's atomic theory
interpreted by de Broglie relation.) Then, very simply (by simple
algebraic equations only) and surprisingly, we obtain such dark
spring characteristics corresponding to basic thermodynamical
characteristics (Bekenstein-Hawking entropy and Hawking
temperature) for corresponding Schwarzschild black hole. Finally,
simple comparison between obtained dark spring characteristics and
Susskind-Horowitz-Polchinski correspondence, a simple
correspondence between strings and dark spring, i.e. classical
linear harmonic oscillator follows. Square root of the elasticity
coefficient of the dark spring corresponds to quotient of one
string coupling g and Newtonian gravitational constant, or,
classical elasticity coefficient of the dark spring corresponds to
reciprocal (inverse) value of the square root of a string state
excitation level N.
\end {abstract}

\section {Introduction}
Within physics two different types of the correct approximation of
an exact theory are possible. First one represents homogeneous
approximation within which all approximate concepts have
equivalent limitation of accuracy in respect to exact theory. For
this reason, within given, self-consistently closed,
approximation, all approximation concepts can be connected simply
or "continuously". An example of such approximation type
represents the classical Newtonian dynamics and gravitation as the
approximation of the Einstein general theory of relativity. Also,
other example of such approximation type represents small
perturbation theory of the quantum mechanics as the exact theory.

Second one approximation type represents heterogeneous or hybrid
approximation within which different approximation concepts can
have different accuracy limitation in respect to exact theory.
Some concepts can be completely exact while some other concepts
can be completely imprecise (even impossible!). For this reason
within given hybrid approximation theory all its characteristic
concepts cannot be connected simply, "continuously", but only
"discretely". An example of such approximation type represents
Bohr atomic theory, i.e. naïve quantum theory in respect to
quantum mechanics as the exact theory. As it is well-known within
Bohr atomic theory electron energies are determined completely
exactly, i.e. identically to quantum-mechanically predictions,
while electron trajectories, used within Bohr atomic theory, do
not exist at all within quantum mechanics. Other example of the
hybrid approximation represents La Place classical mechanical
determination of the dark star radius. As it is well-known [1],
[2] it yields value exactly identical to Schwarzschild, general
relativistic determination of the black hole radius, even if
Laplace used classical expression for light particle kinetic
energy completely inconsistent with general theory of relativity
and even with wave theory of light.

In this work, according to some our previous results [3], [4], we
shall suggest a simplified interpretation of
Susskind-Horowitz-Polchinski correspondence between Schwarzschild
black hole and strings [5], [6]. Firstly, similarly to naïve,
classical mechanical Laplace determination of the Schwarzschild
radius [1], [2], we shall suggest a simple, classical mechanical
equation. It determines amplitude of such sufficiently strong
classical elastic force that forbids escape of a Planck mass
particle moving by speed of light from end of corresponding
classical elastic spring, simply called dark spring. Also, by use
of a formal identity between given elastic force and Schwarzschild
gravitational "force" and according to some our previous results
[3], [4], we shall introduce phenomenologically a simple
quantization rule. It states that circumference (corresponding to
elastic force amplitude equivalent formally to Schwarzschild
radius) holds natural number of corresponding reduced Compton's
wave length. (It is deeply analogous to Bohr's quantization
postulate in Bohr's atomic theory interpreted by de Broglie
relation.) Then, very simply (by simple algebraic equations only)
and surprisingly, we obtain such dark spring characteristics
corresponding to basic thermodynamical characteristics
(Bekenstein-Hawking entropy and Hawking temperature) for
corresponding Schwarzschild black hole. Finally, simple comparison
between obtained dark spring characteristics and
Susskind-Horowitz-Polchinski correspondence, a simple
correspondence between strings and dark spring, i.e. classical
linear harmonic oscillator follows. Square root of the elasticity
coefficient of the dark spring corresponds to quotient of one
string coupling g and Newtonian gravitational constant, or,
classical elasticity coefficient of the dark spring corresponds to
reciprocal (inverse) value of the square root of a string state
excitation level N.

As it is well-known [1], [2] Laplace determined by simple,
classical mechanical method, the radius R of a dark star, i.e. a
star with sufficiently large mass M so that even light cannot
escape from the star surface. Surprisingly given radius is
identical to the Schwarzschild radius of corresponding
Schwarzschild  black hole predicted accurately by general theory
of relativity.

Formally generalizing given Laplace method, suppose that there is
such classical mechanical elastic spring, simply called dark
spring that generates attractive, sufficiently strong classical
elastic force. It forbids escape of a Planck mass particle
$m_{P}=(\frac {\hbar c}{G})^{\frac {1}{2}}$ (where G represents
the Newtonian gravitational constant, $\hbar$ - reduced Planck
constant and c - speed of light) moving by speed of light from end
point of dark spring. Also, suppose that in this case amplitude of
the elastic force is identical to radius R of corresponding
Schwarzschild black hole with mass M. It implies equation
\begin {equation}
         \frac {m_{P}c^{2}}{2} = \frac {kR^{2}}{2} = \frac {Gm_{P}M}{R}
\end {equation}
where k represents dark spring elasticity coefficient. It implies
\begin {equation}
         R = \frac {2MG}{c^{2}} =  \frac {m_{P}^{\frac {1}{2}}c}{k^{\frac {1}{2}}}
\end {equation}
and
\begin {equation}
          M = \frac {Rc^{2}}{2G} = \frac {kR^{3}}{2Gm_{P}} = \frac {m_{P}^{\frac {1}{2}}c^{3}}{2G k^{\frac {1}{2}}} .
\end {equation}

Differentiation of M over k yields
\begin {equation}
          dM = - \frac { m_{P}^{\frac {1}{2}}c^{3}}{4G k^{\frac {3}{2}}} dk
\end {equation}

that implies
\begin {equation}
          dE = d(Mc^{2}) =  -\frac { m_{P}^{\frac {1}{2}}c^{5}}{4G k^{\frac {3}{2}}} dk
\end {equation}

Now we shall shortly repeat formal but very simple calculation
(practically by simple algebraic equations only) of the basic
black hole thermodynamical characteristics, Bekenstein-Hawking
entropy and Hawking temperature, presented in [3], [4]. All this,
as it will be shown, can refer on dark spring too.

Suppose that the mass of the black hole M is quantized and that
given quantums satisfy the following quantization condition
\begin {equation}
      m_{n}cR = n\frac {\hbar}{2\pi}, \hspace{1cm} {\rm for} \hspace{0.5 cm} n = 1, 2,
      ...
\end {equation}
that implies
\begin {equation}
      2\pi R = n\frac {\hbar}{m_{n}c}= n \lambda_{rn} \hspace{1cm} {\rm for} \hspace{0.5 cm} n = 1,
      2,...
\end {equation}
where $\lambda_{rn}=\frac {\hbar}{m_{n}c}$ represents reduced
Compton wave length corresponding to $m_{n}$ for $n=1,2, …$.

Last expression means, in fact, that the circumference of the
black hole horizon  holds n reduced Compton wavelength of the mass
quantums with mass $m_{n}$ for $n = 1, 2, …$.  Obviously, (6), (7)
correspond, in some degree, to remarkable Bohr postulate on the
electron orbital momentum quantization and de Broglie wave
interpretation of this postulate in the atomic physics.

Expression (6) implies
\begin {equation}
       m_{n} = n \frac {\hbar}{2\pi c R} = n m_{1}  \hspace{1cm} {\rm for} \hspace{0.5 cm}  n = 1,
       2,...
\end {equation}
where, according to (2), (3),
\begin {equation}
       m_{1} = \frac {\hbar }{2\pi c R} = \frac {\hbar c}{4\pi G M}=
       \frac {\hbar k^{\frac {1}{2}}} {2\pi m_{P}^{\frac {1}{2}}c^{2}}
\end {equation}
represents the minimal, i.e. ground mass of the mass quantums.

Further, suppose that black hole, i.e. dark spring mass quantums
do a statistical ensemble. In other words, suppose that there is a
gravitational self-interaction of the black hole, i.e. dark spring
which can be described statistically. Suppose that in the
thermodynamical equilibrium almost all quantums occupy ground mass
state. It implies that black hole, i.e. dark spring mass quantums
represent the Bose-Einstein quantum systems, i.e. bosons. Then
Bekenstein-Hawking entropy S can be phenomenologically, according
to (3), (9), determined by
\begin {equation}
  S = k_{B}\frac {M}{m_{1}}= k_{B}\frac {4 \pi G M^{2}}{\hbar c}= k_{B} \frac { \pi m_{P}c^{5}}{G \hbar k}
\end {equation}
where $k_{B}$  represents the Boltzmann constant.

Differentiation of S (10) over M or k yields
\begin {equation}
  dS = k_{B} \frac {8\pi GM }{\hbar c}dM = k_{B} \frac {8\pi GM }{\hbar c^{3}}dE =
  - k_{B} \frac {\pi m_{P}c^{5}}{G\hbar k^{2}}dk       .
\end {equation}

Then, according to the first thermodynamical law $dE=TdS$, (5) and
(8) simply imply Hawking black hole, i.e. dark spring temperature
\begin {equation}
  T  =  \frac {\hbar c^{3}} { k_{B}8\pi GM }=  \frac {\hbar k^{\frac {1}{2}}}{ k_{B}4\pi m^{\frac {1}{2}}_{P}}          .
\end {equation}

Now we shall discuss obtained results. First of all there is a
serious question: holds formally introduced dark spring any real
(within some approximation) physical sense.

Answer on given question can be obtained by comparison of the
presented expressions for black hole characteristics with
corresponding expressions in Susskind -Horowitz-Polchinski
correspondence between Schwarzschild black hole and strings [5],
[6]. It is sufficient to compare our and Susskind
-Horowitz-Polchinski mass and entropy.

In the natural units system ($\hbar=c=k_{B}=1$) and
$m_{B}=G^{\frac {1}{2}}$ expressions (3), (10) turn out in
\begin {equation}
  M = \frac {1}{2G^{\frac {5}{4}}k^{\frac {1}{2}}}
\end {equation}
\begin {equation}
  S = \frac {\pi}{G^{\frac {3}{2}}} \frac {1}{k}  .
\end {equation}
Corresponding Susskind-Horowitz-Polchinski expressions are
\begin {equation}
  M = \frac {1}{(2G) ^{\frac {1}{2}}N^{\frac {1}{2}}}
\end {equation}
\begin {equation}
  S = 2\pi N^{\frac {1}{2}}
\end {equation}
where N represents corresponding string state excitation level
that determines one string coupling by expression
\begin {equation}
   g = \frac {1}{2^{\frac {1}{2}}N^{\frac {1}{4}}}                                    .
\end {equation}
Equivalence between (13) and (16) as well as between (14) and (17)
yields
\begin {equation}
    k = \frac {1}{2G^{\frac {3}{2}} N^{\frac {1}{2}}} = \frac {g^{2}}{G^{\frac {3}{2}}}                        .
\end {equation}

In this way it is proved that dark spring elasticity coefficient
is unambiguously determined by one of two string coupling. Also,
other string coupling $\alpha '$, as it is known [5], [6] is
practically identical to $R^{2}$. It means that dark spring
amplitude R representing square root of other string coupling is
unambiguously determined by this other coupling. For this reason
it can be concluded that string corresponding to Schwarzschild
black hole in Susskind-Horowitz-Polchinski sense can be
simplifiedly effectively presented as a classical linear harmonic
oscillator, i.e. mentioned dark spring.

In conclusion the following can be shortly repeated and pointed
out. In this work, according to some our previous results, we
suggest a simplified interpretation of
Susskind-Horowitz-Polchinski correspondence between Schwarzschild
black hole and strings. Firstly, similarly to naïve, classical
mechanical Laplace determination of the Schwarzschild radius, we
suggest a simple, classical mechanical equation. It determines
amplitude of such sufficiently strong classical elastic force that
forbids escape of a Planck mass particle moving by speed of light
from end of corresponding classical elastic spring, simply called
dark spring. Also, by use of a formal identity between given
elastic force and Schwarzschild gravitational "force" and
according to some our previous results, we shall introduce
phenomenologically a simple quantization rule. It states that
circumference (corresponding to elastic force amplitude equivalent
formally to Schwarzschild radius) holds natural number of
corresponding reduced Compton's wave length. (It is deeply
analogous to Bohr's quantization postulate in Bohr's atomic theory
interpreted by de Broglie relation.) Then, very simply (by simple
algebraic equations only) and surprisingly, we obtain such dark
spring characteristics corresponding to basic thermodynamical
characteristics (Bekenstein-Hawking entropy and Hawking
temperature) for corresponding Schwarzschild black hole. Finally,
simple comparison between obtained dark spring characteristics and
Susskind-Horowitz-Polchinski correspondence, a simple
correspondence between strings and dark spring, i.e. classical
linear harmonic oscillator follows. Square root of the elasticity
coefficient of the dark spring corresponds to quotient of one
string coupling g and Newtonian gravitational constant, or,
classical elasticity coefficient of the dark spring corresponds to
reciprocal (inverse) value of the square root of a string state
excitation level N.

\vspace{1.5cm}

 {\large \bf References}

\begin {itemize}

\item [[1]] P. S. Laplace, {\it Exposition du Systeme du Monde}, Vol. II (Paris, 1796)
\item [[2]] V. Stephani, {\it La Place, Weimar, Schiller and the Birth of Black Hole Theory}, gr-qc/0304087
\item [[3]] V. Pankovic, M. Predojevic, P. Grujic, Serbian Astr. J {\bf 176} (2008) 15, astro-ph/0709.1812
\item [[4]] V. Pankovic, {\it Black Holes - A Simplified Theory for Quantum Gravity Non-Specialists}, gen-ph/0911.1026
\item [[5]] L. Susskind, {\it Some Speculations about Black Hole Entropy in String Theory} hap-th/9309145
\item [[6]] G. T. Horowitz, J. Polchinski, {\it A Correspondence Principle for Black Holes and Strings}, hep-th/9612146

\end {itemize}

\end {document}